# Deep learning the high variability and randomness inside multimode fibres

*Pengfei Fan, Tianrui Zhao & Lei Su*[*]

*School of Engineering and Materials Science, Queen Mary University of London, UK*
l.su@qmul.ac.uk

Multimode fibres (MMF) are remarkable high-capacity information channels owing to the large number of transmitting fibre modes[1-3], and have recently attracted significant renewed interest in applications such as optical communication, imaging, and optical trapping[4-15]. At the same time, the optical transmitting modes inside MMFs are highly sensitive to external perturbations and environmental changes, resulting in MMF transmission channels being highly variable and random[16-18]. This largely limits the practical application of MMFs and hinders the full exploitation of their information capacity. Despite great research efforts made to overcome the high variability and randomness inside MMFs, any geometric change to the MMF leads to completely different transmission matrices, which unavoidably fails at the information recovery. Here, we show the successful binary image transmission using deep learning through a single MMF, which is stationary or subject to dynamic shape variations. We found that a single convolutional neural network has excellent generalisation capability with various MMF transmission states. This deep neural network can be trained by multiple MMF transmission states to accurately predict unknown information at the other end of the MMF at any of these states, without knowing which state is present. Our results demonstrate that deep learning is a promising solution to address the variability and randomness challenge of MMF based information channels. This deep-learning approach is the starting point of developing future high-capacity MMF optical systems and devices, and is applicable to optical systems concerning other diffusing media.

We use a simple digital micromirror device (DMD) based single MMF transmission system[19] to demonstrate the idea, as shown in Fig. 1(a). The DMD can be controlled to display any binary pattern by turning 'ON' and 'OFF' individual micromirrors, simulating a 2D object imaging from the DMD through the MMF to the camera. The experimental setup details are given in Methods. As the first step to verify our idea, we use transmission matrix (TM) calculated output speckle images for deep neural network training and prediction[20]. Firstly we calculate the MMF TM using 8000 input-output pairs collected from the experimental system. The measured TM is verified experimentally to have an average accuracy of 98% for output speckle pattern calculation (details in Methods). We then use the processed binary images of handwritten digits downloaded from the MNIST handwritten digits database[21,22] as the DMD input pattern. The binary processing is to convert the handwritten digit into 36×36 pixel image format that directly matches the fibre mode count and eigenchannels for optimal delivery[19] (details in Methods). As shown in Fig. 1(b), the MMF output speckle images are calculated using TM and the binary handwritten digits. We calculate 7,040 speckle images corresponding to different input binary images for the digits '0-9' (704 speckle images for each digit). Out of these speckle images, 5,632 speckle images and their corresponding binary DMD input patterns are randomly selected as our training image dataset to train a deep convolutional neural network (CNN) (details in Supplementary Information), as shown in Fig. 1(c). This training procedure only needs to be performed once, and the CNN is then fixed. The remaining 1,408 speckle images and their corresponding binary DMD input patterns were used for testing the final network model (Fig. 1(c)). To further quantify the performance of the final trained CNN, we calculate another 780 output speckle images for handwritten letters downloaded from a different EMNIST Letters database[23].

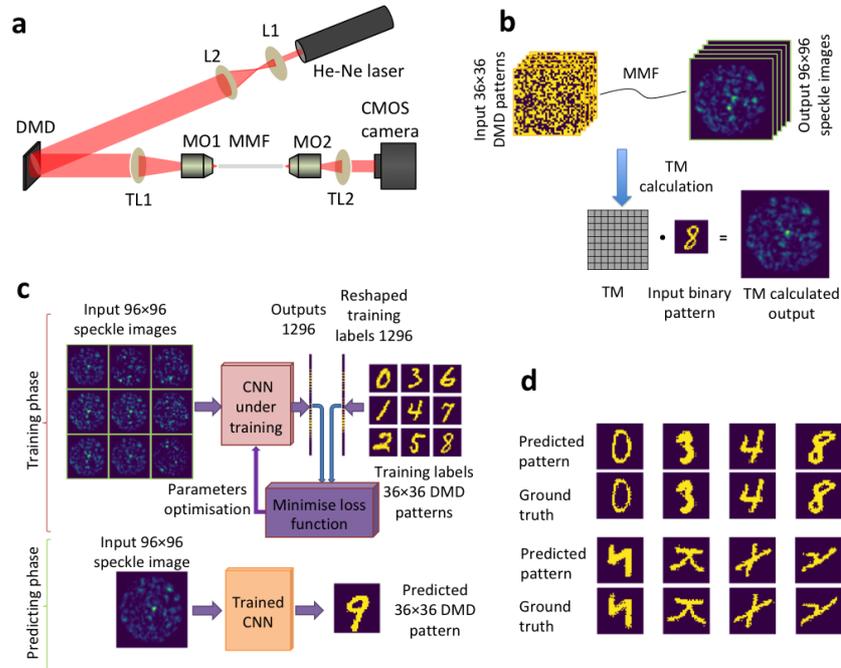

Fig. 1 Input DMD pattern prediction using CNN with TM calculated MMF output speckle images. **a**. Experimental setup (details in methods). **b**. MMF TM calculation using experimentally measured input-output pairs. **c**. Schematic to show CNN training and prediction. **d**. CNN prediction results for both hand-written digits and letters not used in network training.

Examples of the prediction results are shown in Fig. 1 (d) (more results in Supplementary Information). The prediction accuracy, defined as the percentage of correctly predicted pixels, is used to assess the prediction performance. For the 1,408 testing speckle images not used in the training process, the average prediction accuracy between the predicted binary images and the ground truth is 98.74%. For 780 handwritten letters testing images from the other database, the network output demonstrates 95.22% average accuracy over 780 test images. The results demonstrate that the '0-9' digits trained CNN can be used to predict letter inputs with significantly different shapes.

In the second step, we use the experimentally obtained output speckle images to verify the neural network prediction performance. We experimentally collect input and output pairs to train the CNN instead of using the TM calculated output. Using the experimental configuration shown in Fig. 1 (a), we modulate the DMD input pattern using the binary MNIST handwritten digits database and the EMNIST Letters database, and collect corresponding output speckle images with the CMOS camera. The above input-output-pair collection is repeated for three different geometric states of the MMF, named as G1, G2 and G3 (Fig. 2 (a)). The significant difference in light transmission between these three states is ascertained by evaluating the correlation coefficients of output speckles collected at G1, G2 and G3 for the same input DMD pattern. As shown in Fig. 2 (a), the correlation coefficients between any of these states are below 60%.

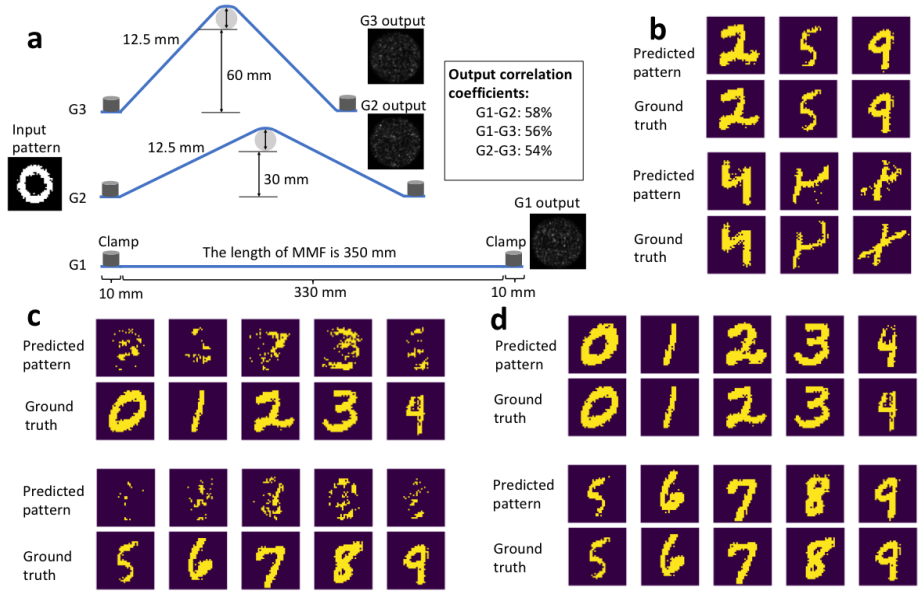

Fig. 2 CNN prediction results at three different MMF geometric states. **a**. MMF configurations: G1, G2 and G3 are three geometric states for the MMF under test (50μm-core, 350mm-long, FG050UGA, Thorlabs). The correlation coefficients between outputs from the same input at G1, G2 and G3 demonstrate the significant difference in transmission properties at these three states. **b**. Examples of prediction results of using G3-trained CNN to predict G3 testing data. **c**. Examples of prediction results of using G1&G2-trained CNN to predict G3 testing data. **d**. Examples of prediction results of using G1&G2&G3-trained CNN to predict G3 testing data. The testing data are not in the training dataset.

To train and test the deep CNN, we experimentally collect three different output speckle pattern datasets at these three different MMF geometries, G1, G2 and G3 respectively. For these geometries, the same DMD input dataset, including approximately 40000 unique '0-9' MNIST handwritten digits and 780 unique EMNIST letters, is used. We use approximately 90% of the experimentally collected input-output pairs from the handwritten digits database to train the deep CNN, and the remaining 10% handwritten digits input-output pairs and all handwritten letter input-output pairs for testing. Firstly, we separately train and test G1, G2 and G3 using their own datasets. The prediction results by using the CNN trained by the data collected under the same MMF geometric states are excellent. Prediction examples for G3 are shown in Fig. 2 (b), with an average accuracy of 97.06% over 3622 testing handwritten digit speckle images, and an average accuracy of 92.32% for 780 handwritten letters from the different EMNIST letter database (extended results in Supplementary Information). This proves our simulation results in Fig. 1 (d) experimentally. The trained CNN can successfully predict digits and letters significantly different in shape under the same MMF geometric state (i.e. the same TM). However, as expected, the trained CNN at one MMF geometric state fails to predict the corresponding inputs of output speckle images at other states. Fig. 2 (c) is an example using G1&G2-trained CNN to predicate the input pattern based on the speckle images collected at G3 and the predicted pattern significantly mismatches the ground truth, where the average prediction accuracy is 86.27% over 3622 testing handwritten figures. This accuracy is over 10% worse than that achieved by G3-trained CNN. Since all the input patterns have large proportions of black edges, this 10% difference means significant mismatching at the centre areas of input patterns.

The following experiment leads to a key finding of our work. We use the combination of data collected at G1, G2 and G3 states to train the CNN. Subsequently, the newly trained CNN is used to predict the input patterns for both the digit and letter testing speckle images collected at the G1, G2 and G3 states, respectively. As depicted in Fig. 2 (d), our mixed G1&G&2G3-trained CNN provides outstanding prediction performance for testing inputs measured at G1, G2 and G3 states. The average prediction accuracies for G1&G2&G3-trained CNN are: 96.96% for 4097 G1 testing data, 96.31% for 3801 G2 testing data, and 96.05% for 3622 G3 testing data. Although slightly lower, these are in accordance with the prediction accuracies achieved by individually trained CNNs for G1, G2 and G3 (97.88%, 97.36%, and 97.06%, respectively) and are highly accurate predictions. The results demonstrate that the proposed deep neural network not only has great performance to predict the input patterns at a certain MMF geometric state, but also exhibits a significant generalisation capacity for different MMF geometric states (i.e. different TMs).

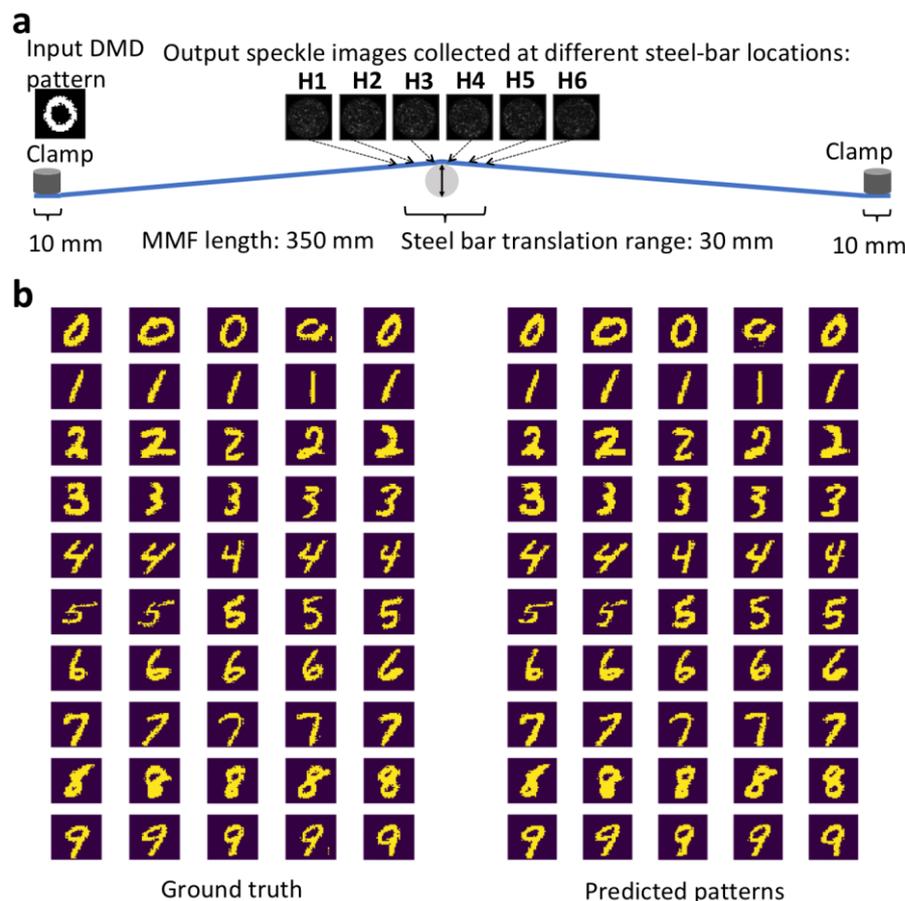

Fig. 3 CNN prediction results when the MMF is subject to continuous shape variations manually applied at a random speed. The output speckle images for the same input DMD pattern are shown for H1-H6. **a**. Experimental configuration to apply continuous shape variations to the MMF under testing. **b**. CNN prediction results.

Inspired by the results above, we design an experiment to test if a deep CNN can be trained to predict input DMD patterns, using output speckle images captured when the MMF is subject to continuous shape variations applied by human at a random speed. Using the experimental setup in Fig. 1(a), we insert a steel bar (diameter 12.5 mm, length 200 mm) under the MMF. The steel bar can be translated within a range of 30 mm in the middle of the MMF, as illustrated in Fig. 3 (a). We ensure that the

MMF under test is in tight contact with the steel bar, so that the strain applied to the MMF by moving the steel bar is strong enough for significant TM variations[14]. At the same time, the strain on the MMF cannot be too strong; otherwise it may overcome the clamping friction, resulting in the fibre-end movements and misalignment. Throughout the data collection process (approximately 110 minutes, including both data uploading time and input-output pair measurement time), the steel bar is manually translated along both MMF axial directions within the 30mm range at a random speed within 0-30mm/s. This ensures continuous and random-speed shape changes applied to the MMF, simulating the situation of manual MMF manipulations. To show significant MMF transmission property variations during the steel-bar translation, we calculate correlation coefficients of the MMF output speckle images at the same input DMD pattern between six different steel-bar locations, H1-H6, which are equally distanced along the 30mm translation range with approximately 5mm between any two adjacent locations. We show in Fig.3 (a) that the output speckle images for the same input DMD pattern at H1-H6 are significantly different from each other with calculated correlation coefficients around 50% (details in Supplementary Information). Note that for our experiment system, the strain required to reach a correlation coefficient lower than 45% is likely to overcome the clamping friction and to disrupt the alignment at MMF ends.

We acquire 39641 speckle images for input DMD patterns modulated with different '0-9' MNIST handwritten digits. Out of these images, 35676 speckle input images and their corresponding binary DMD labels were randomly selected to be used as our training dataset, and the remaining 3965 speckle inputs and their corresponding patterns formed our test images to blindly quantify the average performance of the network. It is worth noting that the handwritten digit input dataset is only used once for all the shapes incurred during the steel-bar translation over the 30mm MMF. This is different from the three different MMF geometric states experiment in Fig. 2, where the whole input dataset is repeatedly used for three different states. This increases the difficulty in predictions, as there are less training data and all input patterns in the handwritten dataset are different from each other. Fig. 3 (b) illustrates the prediction results of CNN trained by data collected when the MMF is subject to continuous shape variations. It is clear that the shapes, tilting angles and pixel-level details of the predicted digits resemble the actual inputs very well. The average prediction accuracy is 96.48% over 3965 testing images.

Our findings suggest that the high variability and randomness inside MMFs can be overcome by training a deep neural network with all the possible variations that may occur to a certain MMF based optical system. Particularly, we demonstrate experimentally that distorted images (DMD input patterns) through a MMF subject to continuous shape variations can be recovered successfully by an appropriately trained deep CNN without knowing the exact MMF geometrical state or the TM. Our work paves the way for fully exploring the high-information capacity of MMF channels, and will lead to future new MMF-based optical systems for applications such as imaging and communications.

## Methods:

**Experimental Setup**
The laser beam (632.8nm, 17mW, 25-LHP-925-230, Melles Griot) is expanded by Lens 1 (L1, Mounted Rochester Aspheric Lens, focal length = 11.00 mm, NA = 0.30, A397TM-A, Thorlabs) and Lens 2 (L2, Bi-Convex Lens, focal length = 100.0 mm, LB1187-A-ML,Thorlabs) and is then projected onto a digital micromirror device (DMD, 1024×768 micro-mirrors, mirror size 13.68μm×13.68μm, mirror tilt +/- 12 degrees, operating at 20Hz, maximum frequency 9.8kHz, Discovery 1100, Texas Instruments). Driven by an interface board (ALP-1, ViALUX), the DMD can display any arbitrary binary pattern by turning 'ON' and 'OFF' individual DMD micromirrors. The incident laser beam can be modulated by the displayed DMD pattern and is subsequently coupled into the proximal end of a multimode fibre (MMF, 50μm-core, 35cm-long, FG050UGA, Thorlabs) through a tube lens (TL1, AC254-200-A-ML, focal length = 200 mm, Thorlabs) and a microscope objective (MO1 Nikon CFI Plan Achro 40×, 0.65 NA, 0.56 mm WD, tube lens focal length: 200 mm). At the other end of the MMF, a microscope objective (MO2, Olympus 20× Plan Achro, 0.4 NA, 1.2 mm working distance, tube lens focal length: 180 mm) and a tube lens (TL2, AC254-200-A-ML, focal length = 200 mm, Thorlabs) were used to expand the output beam from the MMF distal end. The output was then sent to a CMOS camera (2048×2048 pixels, 6.5 μm x 6.5μm pixel sizes, operating at 20fps, maximum frame rate 100fps, C1140-22CU, Flash4.0, Hamamatsu).

**Multimode fibre transmission matrix measurement**
In Fig. 1 (a), we set 2×2 micromirror pixels as a single macro pixel in the modulation to enhance the difference between the "ON" and the "OFF" pixels. The input DMD modulation pattern consists of 36×36 (N = 1296) macro pixels and the output pixels captured by CMOS camera is 96×96 (M = 9216). N is determined by the maximum number of modes supported in the MMF to ensure sufficient degrees of freedom. M is determined by the MMF core diameter, NA and the magnification of the objective lens. We generated P = 8000 random binary patterned input fields, with an "ON" to "OFF" pixel ratio of 50:50. The response of a single macro pixel was therefore measured P/2 times over the cycle. The DMD is operated at 250 Hz. The system stability is monitored by calculating correlation coefficients of outputs for the same input over time and is above 99%. The input and output data was processed and the TM was calculated with the prVBEM (phase retrieval Variational Bayes Expectation Maximization) algorithm[19,24], which requires $M/N \geq 2$ in order to retrieve phase information successfully.

The performance of the TM is evaluated with the correlation coefficient (corr2/corr functions in MATLAB) between TM calculated outputs and experimentally measured outputs at the same input DMD patterns (random binary pattern at 50:50 "ON" to "OFF" ratio). By evaluating over 8000 input DMD patterns, we obtain an average correlation coefficient of over 98% for the TM used in our simulation.

**Binary DMD input pattern preparation**
To initially prepare the raw label training data for the deep neural network, we downloaded images of handwritten digits from the MNIST handwritten digits database[21]. The raw images consisting of 28 × 28 pixels are converted to images with 36×36 pixels to match the input DMD pattern requirement. Since the on-board rotation of micromirrors can only realise binary intensity modulation, the 36×36 pixel images are then converted into binary images (implemented using *im.convert* function within the Python Imaging Library) to yield the final digits training labels for the deep neural network. Since we use 2×2 DMD pixels as a micromirror pixel, the binary images with 36×36 pixels are used to modulate 72×72 pixels on the DMD.

## Implementation
Keras library version 2.1.5 with Python version 3.6.4 was used to implement the program. The architecture of the deep neural network was set up by using TensorFlow framework back-end version 1.2.1 (Google). We utilised Queen Mary's Apocrita HPC facility to train our network's models. The training phase of the network was performed under the Linux Singularity container using a Tesla K80 GPU card (Nvidia) and a 16 Core Xeon E5-2683V3 processor (Intel) with 7.5GB RAM requested for each core. For the network's training stage, ~10 times speedup performance was implemented by GPU acceleration, compared to a single CPU for network training. In the prediction phase, a laptop computer is used (Intel Core i7-7600K CPU @ 2.8GHz, 8GB of RAM, running a Microsoft Windows 10 professional operating system). The computing time for training and prediction for different datasets are available in Supplementary Information.

## Data availability
All data generated and analysed during this study is available from the corresponding author on reasonable request.

## Affiliations
School of Engineering and Materials Science, Queen Mary University of London, UK

## Contributions
P.F. and T.Z. contributed equally to this work. L.S. conceived the idea and designed the experiment. T.Z. set up and performed the experiments. P.F. designed the neural network and performed data analysis, and all authors analysed the results. L.S. guided the work, and wrote the manuscript with contributions from all authors.

## Acknowledgement
This work was supported by Engineering and Physical Sciences Research Council (grant number EP/L022559/1 and EP/L022559/2); Royal Society (grant numbers RG130230 and IE161214). This research utilised Queen Mary's Apocrita HPC facility, supported by QMUL Research-IT. http://doi.org/10.5281/zenodo.438045

## Competing interests
The authors declare no competing interests.

# Supplementary Information

**Deep convolutional neural network Architecture**
The detailed architecture for the proposed deep neural network is depicted in Fig. S1. To infer the relationship between the output multimode fibre (MMF) speckle images and the modulated digital micromirror device (DMD) patterns, the feature map dimension (i.e., the number of channels) used for the three convolutional layers is 5, 10 and 24 respectively, as illustrated in Fig. S1. This is empirically determined to balance the trade-off between the network complexity and prediction output time. The size of kernels (filters) used throughout the network's convolutional layers is 3×3 elements. The input image is mapped into five output feature maps by the first convolutional layer, which is followed by a batch normalization (BN) layer[1] and a dropout layer[2] (detailed below in this section). The first convolutional layer maps the original 96×96 pixels speckle image into five 94×94-pixel feature maps. The second convolutional layer maps these five 94×94-pixel feature maps to ten 46×46-pixel output feature maps, as detailed in Fig. S1. This approximately quarters the size of the feature maps by utilising a two-pixel stride convolution, while a single pixel stride convolution is performed in the other two convolutional layers in the network. The third convolutional layer maps these ten 46×46 pixels feature maps into 24 44×44-pixel feature maps.

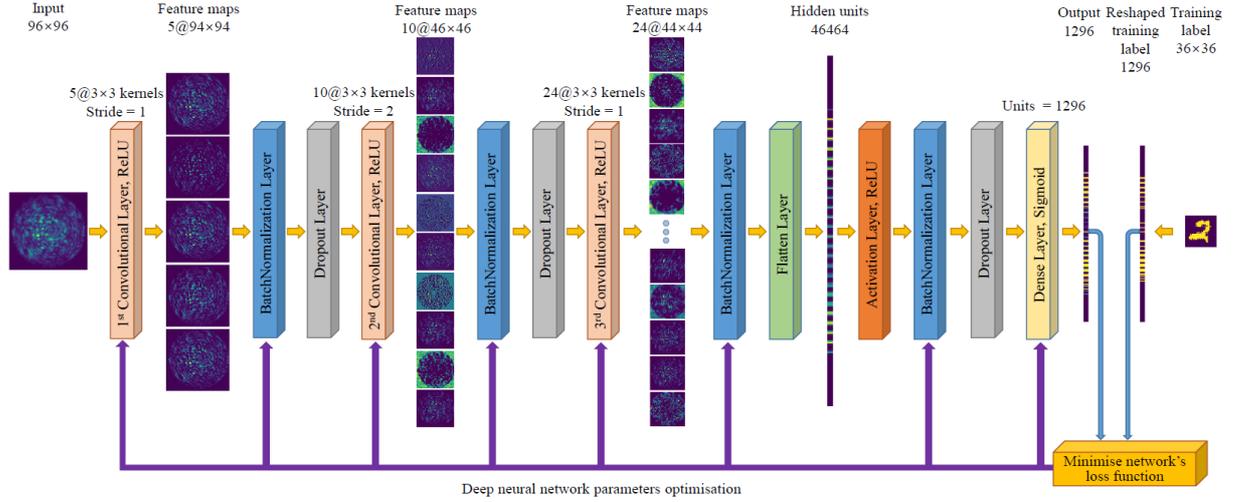

Fig. S1 The proposed deep neural network architecture

Each convolutional layer consists of a convolution operation (determined by the learnable kernels and their biases) and an activation function (here ReLU is used), as shown in Fig. S1. The ReLU is an activation function and performs the calculation ReLU(x) = max(0, x). The mathematic expression for each convolutional layer can be summarised as:

$$X_{out} = \text{ReLU}(X_{in} * W + B), \qquad (S1)$$

where $X_{in}$ refers to the input to the convolutional layer, $*$ denotes the convolution operation, $W$ denotes an ensemble of learnable convolution kernels, $B$ refers to the bias, $X_{out}$ is the output from the convolutional layer. The output feature maps from the convolutional layer in the network are expressed as:

$$g_i = \sum_j f_j * w_{i,j} + \beta_i \Omega, \qquad (S2)$$

where $w_{i,j}$ is a learnable 2D kernel (i.e., the (*i*,*j*)-th kernel of $W$) applied to the *j*-th input feature map $f_j$, $\beta_i$ is a learnable bias term, $g_i$ is the *i*-th *M*×*M*-pixel output feature map, and $\Omega$ is an *M*×*M*-pixel all-ones matrix. Note that the maximum values of *i* and *j* correspond to the numbers of the output and input feature maps respectively.

The three convolutional layers allow better data representations by extracting multiple high-level features from images. Theses extracted high-level features can be combined cheaply by adding a fully-connected layer to learn a non-linear function in the feature space effectively. To be specific, another BN layer is followed by a flatten layer. This flatten layer yields 24 × 44 × 44 = 46464 one-dimensional flatten units, which are subsequently activated by a ReLU activation layer. After the last BN layer and the last dropout layer, we utilised a dense layer (fully connecting the neighboring layers by a weighting value), which makes the model end-to-end trainable. The activation function used in the dense layer of our deep neural network is a sigmoid function. It has a characteristic 'S'-shaped curve bounding the output to the 0-1 range (which matches our binary labels retrieval task), defined by the formula:

$$S(x) = \frac{1}{1+e^{-x}} = \frac{e^x}{e^x+1}, \quad (S3)$$

BN layers and dropout layers are added as elemental building blocks in our deep neural network to enhance the network performance. BN layers accelerate the learning convergence, and dropout layers prevent the neural network from overfitting.

A BN layer conducts an affine transformation with the following equation:

$$y = \gamma x + \beta, \quad (S4)$$

where $\gamma$ and $\beta$ are a pair of parameters learnt for each activation in feature maps to scale and shift the normalized value. The experimental results of the four structures provided in Fig. S2 show that the BN layers enhance the network performance.

To avoid overfitting, dropout layers are essential in the building blocks of the network. In our experiment, the dropout rate was empirically determined as 0.4 for good performance. The results in Fig. S2 suggest that removing the dropout layers deteriorates the network performance significantly.

The deep neural network architecture discussed-above provides three major benefits: first, the inverse problem becomes a learnable operation with supervised learning; second, the fewer convolutional operations enable outstanding representation power of the network; and third, the appropriate use of BN layers and dropout layers maximises the capability of the network, speeds up the convergence and prevents the network from overfitting. All these lead to the excellent performance of the deep neural network and its significant generalisation to the complex light propagation process inside multimode fibres.

**Data pre-processing**
For experimentally collected speckle-image training samples, the intensity of some transmitted light would inevitably overstep the dynamic range of the camera sensor resulting in bad data for the training and testing phase of the deep neural network. Bad data elimination is performed on each group dataset to further refine the data validity. Here, images are removed from the dataset if they fall into the following two categories: i) No pixel in the image has an intensity value greater than 10,000; or ii) one or more pixels in the image have intensity values greater than 65,000. Thresholds for data elimination were empirically determined by the intensity dynamic range (0-65,535) of the camera (C1140-22CU, Flash4.0, Hamamatsu) to provide the optimal balance between high learning quality and full information preservation. The percentage of removed bad data is usually less than 5% of the collected dataset.

After data pre-processing, this intensity-only speckle image consisting of 96×96 pixels and its corresponding binary pattern with 36×36 pixels (reshaped to 1,296 pixels), form an input-label pair, which is used for the network's training and testing.

**Network training**

The designed network architecture determines the total number of its parameters. Network training phase is to optimise these parameters. The details (including parameters) of our deep neural network are provided in Table S1. The total number of parameters in the network is 60,407,346, consisting of 60,314,340 trainable params and 93,006 non-trainable params. The non-trainable weights, namely the mean and the variance vectors, are maintained by BN layers and updated via layer updates but not through the back propagation[1].

Table S1. Details of the deep neural network

| Layer (type) | Output Shape | Param # |
|---|---|---|
| Conv2D_1 | (94, 94, 5) | 50 |
| BN_1 | (94, 94, 5) | 20 |
| Dropout_1 | (94, 94, 5) | 0 |
| Conv2D_2 | (46, 46, 10) | 460 |
| BN_2 | (46, 46, 10) | 40 |
| Dropout_2 | (46, 46, 10) | 0 |
| Conv2D_3 | (44, 44, 24) | 2184 |
| BN_3 | (44, 44, 24) | 96 |
| Flatten_1 | 46464 | 0 |
| Activation_1 | 46464 | 0 |
| BN_4 | 46464 | 185856 |
| Dropout_3 | 46464 | 0 |
| Dense_1 | 1296 | 60218640 |
| Total params: 60,407,346 | | |
| Trainable params: 60,314,340 | | |
| Non-trainable params: 93,006 | | |

As shown in Fig. S1, we use sigmoid as the activation function to predict the final binary patterns in the output layer of the network. In fact, it is useful to consider a sigmoid output layer with Binary Cross Entropy cost rather than other quadratic cost functions (such as MSE). It has been proved that the quadratic cost functions will slow down the training phase during error backward propagation, while cross entropy overcomes this issue by dynamically controlling the speed of learning with the output changing errors[3].

The Binary Cross Entropy cost function, however, led to disappointing performance in a preliminary experiment as a result of the imbalanced classification problem[4], where the classes of the data are not represented equally. In our case, the handwritten digits dataset is an imbalanced dataset and the ratio of positive targets (i.e., '1's in the binary pattern) to all pixels is only 10% to 20%. That is to say, one handwritten digit just occupies a small part of the entire 36×36 pixels. The initial model is highly possible predicting negative targets regardless of the data. Therefore, we introduce adaptively weighted binary cross entropy as the loss function for training the network to deal with highly imbalanced data.

The usual cross-entropy cost can be expressed as[5]:
$$t*-log(sigmoid(y))+(1-t)*-log(1-sigmoid(y)), \quad (S5)$$
where $t$ is the expected targets (labels), $y$ denotes the logits of the predicted outputs. In implementation, targets and logits must have the same type and shape.

We use a value called *pos_weight*, which tunes the trade-off between the two metrics for classification tasks, namely 'recall' and 'precision', by up- or down-weighting the cost of a positive error relative to a negative error. This can be seen from the fact that *pos_weight* is introduced as a multiplicative coefficient for the positive targets term in the loss expression:

$$t*-log(sigmoid(y))*pos\_weight+(1-t)*-log(1-sigmoid(y)),$$ (S6)

The argument *pos_weight* needs to be tuned. For example, if we have the ratio of positives to negatives, p/n, we set *pos_weight* to be n/p, such that both categories contribute equally in total. In our implementation, the mean value of the ratio in all binary patterns from each experiment was calculated separately according to the datasets, and was assigned to pos_weight. The rebalanced classes provide a rational basis to determine the threshold of binary processing. This threshold used in our implementation is 0.5.

Following network's loss function minimisation, the error between the output pattern and its corresponding ground truth is backpropagated through the network, and a stochastic optimisation method called the AdaDelta optimization[6] is used to optimise the network's parameters. In our implementation, the learning rate parameter is empirically set as 0.1, and the total batch is split to mini-batches with 32 patches each. The entries of kernels (with $3\times 3$ elements) used in convolutional layers are initialised by using glorot_uniform[7], specifically,

$$w_{i,j} \sim \text{Uniform}(-\frac{\sqrt{6}}{\sqrt{n_{in}+n_{out}}}, \frac{\sqrt{6}}{\sqrt{n_{in}+n_{out}}}),$$ (S7)

where $n_{in}$ and $n_{out}$ are the number of input and output channels, respectively. All the bias terms (for instance $\beta_j$) are initialised as 0.

As an example, the training and validating losses of CNNs with different structures trained by data collected when the MMF is subject to continuous shape variations (CSV) are illustrated in Fig. S2. After 200 epochs, the error between the network outputs and the binary handwritten digits drops to 0.1 and converges steadily, depicted in Fig. S2 (a). This suggests that the network is trained sufficiently. The testing results of this CSV-trained CNN are shown in Fig. 3 (b) of the main text. In contrast, the results are completely unsatisfying without the dropout or BN layers, as shown in Fig. S2 (b)-(c). This demonstrates the importance of BN and dropout layers in our CNN.

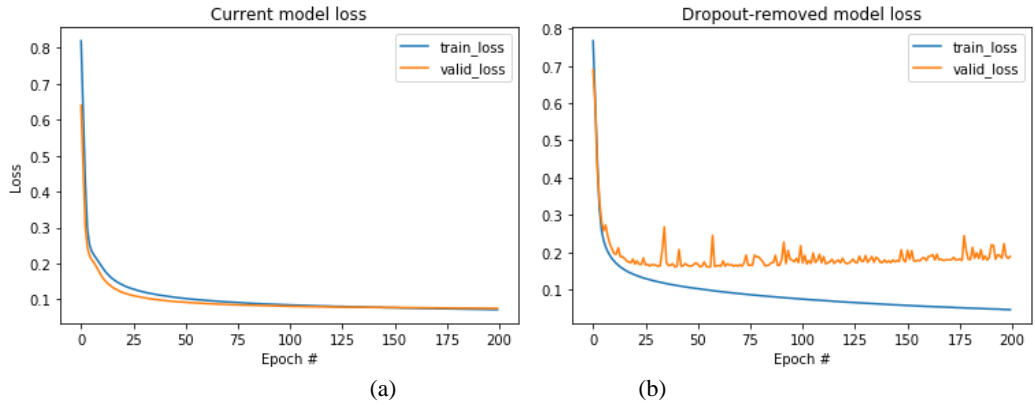

(a)          (b)

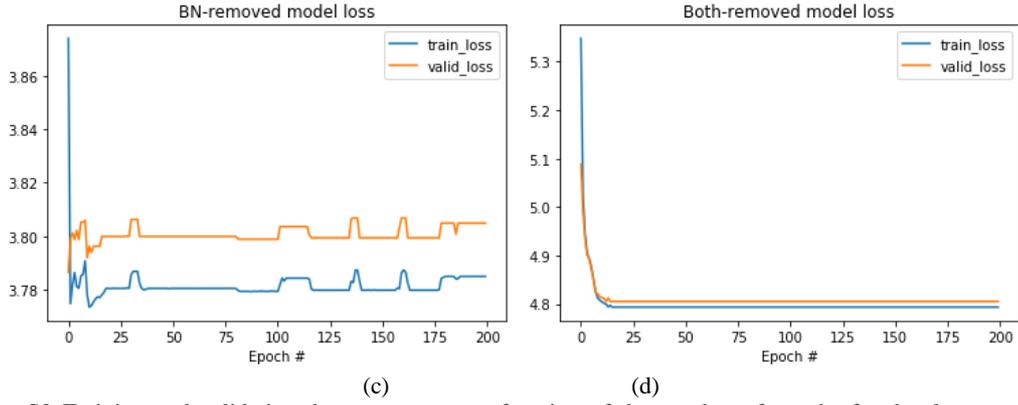

(c)                      (d)

Fig. S2 Training and validation dataset errors as a function of the number of epochs for the deep neural network model in comparison with other three structures. (a) loss curve of the proposed network's model; (b) loss curve with the dropout layers removed network's model; (c) loss curve with the BN layers removed network's model; and (d) loss curve with both BN and dropout layers removed network's model.

The training time of the deep neural networks for different image datasets mentioned in the Main Text is summarised in Table S2 (with the implementation configuration detailed in Implementation in the Main Text).

Table S2. Deep neural network training details for different datasets.

| Models | Number of input-output training patches | Validation set | Training time per epoch (*sec*) | Total training time (200 epoch) |
|---|---|---|---|---|
| **TM-trained CNN** | 4,224 patches | 1,408 images | 16 | 52min |
| **G3-trained CNN** | 24,444 patches | 8,148 images | 92 | 5hr,6min |
| **CSV-trained CNN** | 26,757 patches | 8,919 images | 101 | 5hr,36min |

**Network testing**

The sufficiently trained CNN is capable of predicting the input DMD pattern, by generating a 36 × 36-pixel binary image based on a 96 × 96-pixel speckle image captured at the camera.

**Testing results for transmission matrix (TM) generated data**

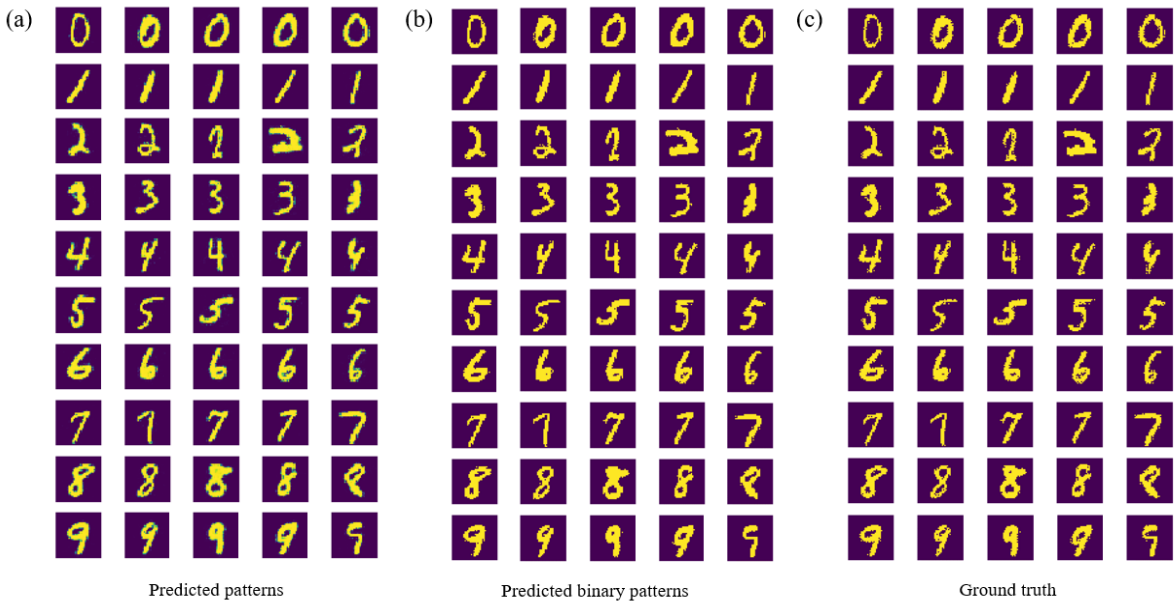

Fig. S3 Extended TM-trained CNN output images corresponding to Fig. 1(d) in the Main Text. (a) Predicted patterns; (b) Predicted binary patterns; (c) Comparison images of ground truth patterned on the DMD.

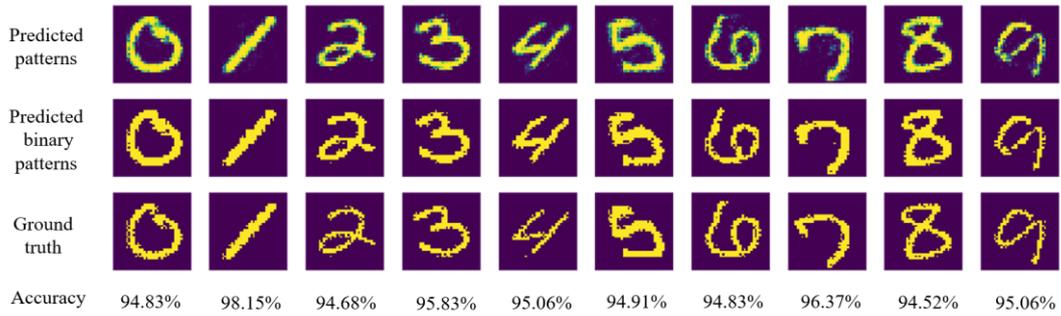
Fig. S4 Extended TM-trained CNN output images with worst performance

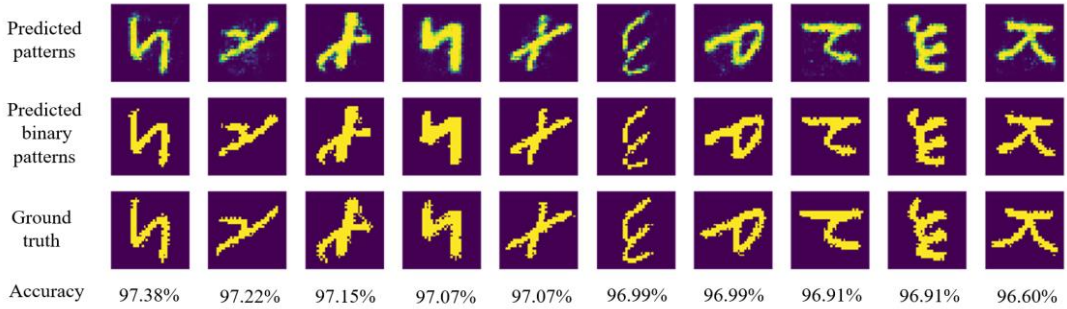
Fig. S5 Extended TM-trained CNN output images of predicted letters

**Testing results for data collected at three different MMF geometries: G1, G2 and G3**

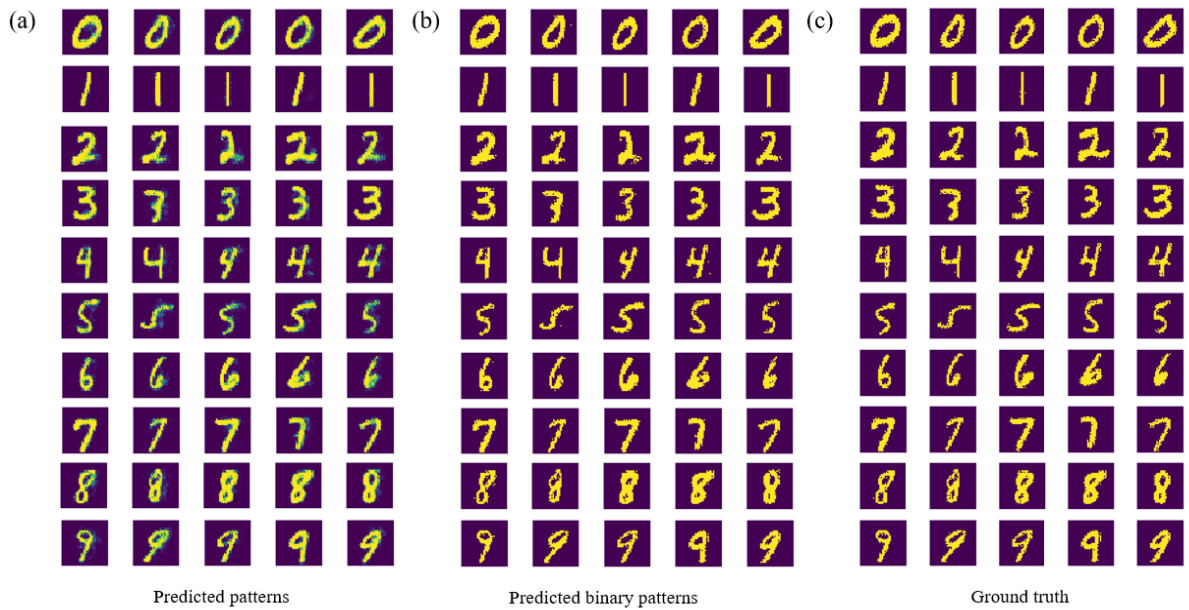
Fig. S6 Extended G3-trained CNN output images. (a) Predicted patterns; (b) Predicted binary patterns; (c) Comparison images of ground truth patterned on the DMD.

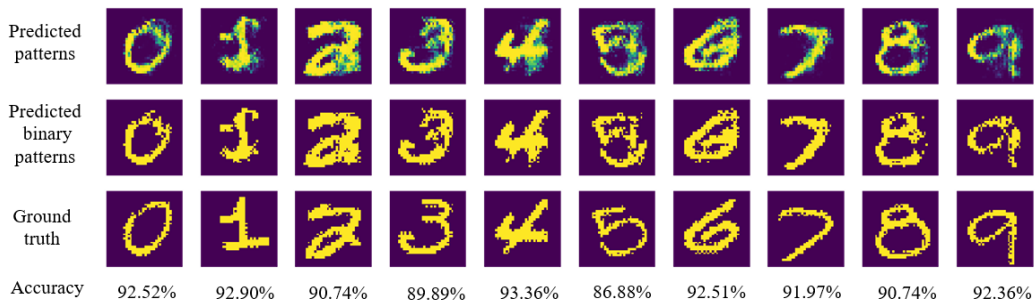
Fig. S7 Extended G3-trained CNN output images with worst performance

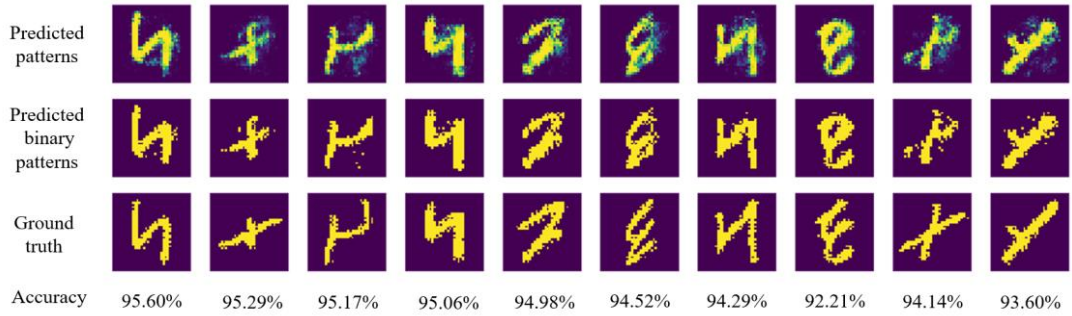

Fig. S8 Extended C3-trained CNN output images of predicted letters

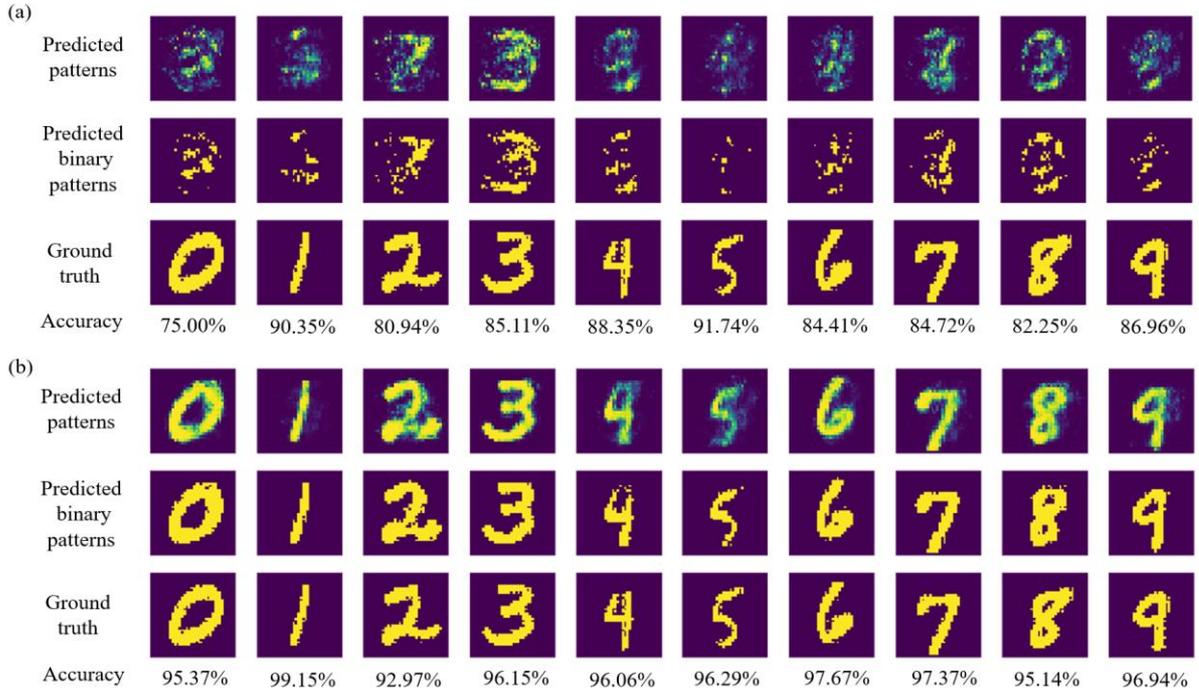

Fig. S9 (a) The output images of G1G2-trained CNN on the G3 test dataset; (b) The output images of mixed G1&G2&G3-trained CNN on the G3 test dataset. Note that in (a) even at accuracy over 90%, the shape of the predicted result is still far from its ground truth. Using the digit '5' as an example, the predicted '5' in (a) is completely unrecognisable at 91.74% accuracy, compared to the predicted result in (b).

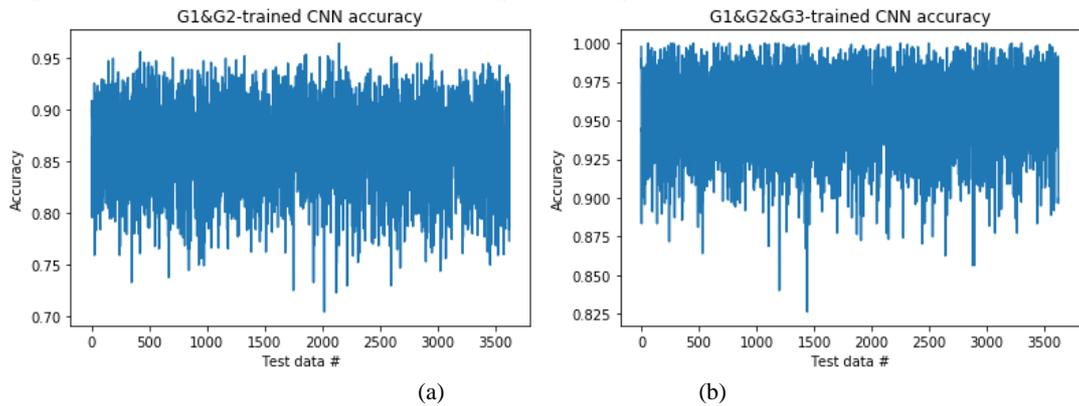

Fig. S10 Prediction accuracy distribution of different CNNs on the G3 test dataset. (a) G1&G2-trained CNN accuracy on the G3 test dataset; (b) G1&G2&G3-trained CNN accuracy on G3 test dataset

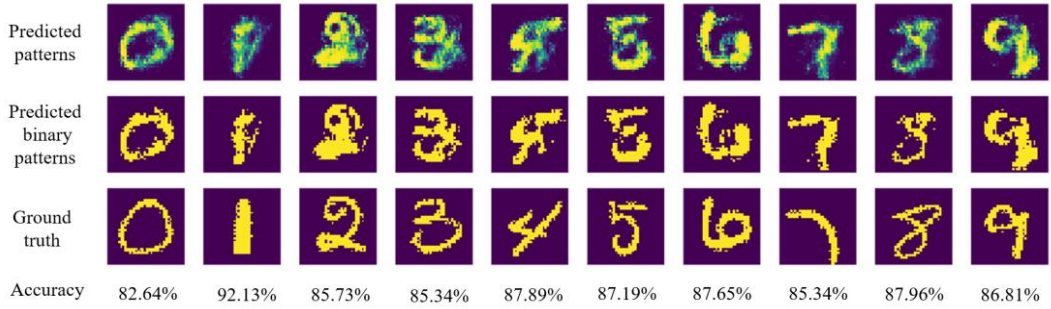

Fig. S11 Extended G1&G2&G3-trained CNN output images on the G3 test dataset with worst performance corresponding to Fig. S9 (b) and Fig. 2 (d) in the Main Text

**Testing results with the MMF subject to continuous shape variation (CSV)**

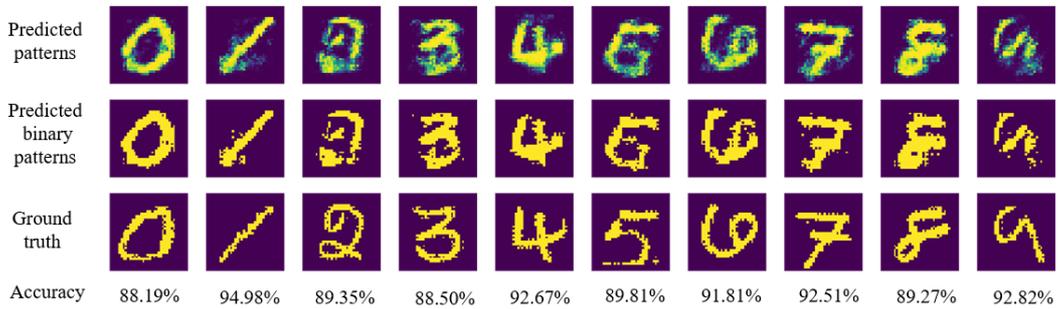

Fig. S12 Extended CSV-trained CNN output images with worst performance corresponding to Fig. 3 (b) in the Main Text

**Network performance evaluation:**

Testing images were used to numerically quantify the performance of our trained CNN models. From the perspective of classification evaluation, the output image of the network is quantified by using the accuracy and the f1_score, which are calculated between the predicted binary pattern and the ground truth. On the other hand, the mean squared error (MSE), the correlation coefficient (corr) and the structural similarity index (SSIM)[8] are also calculated between the predicted pattern and the ground truth from the perspective of regression evaluation.

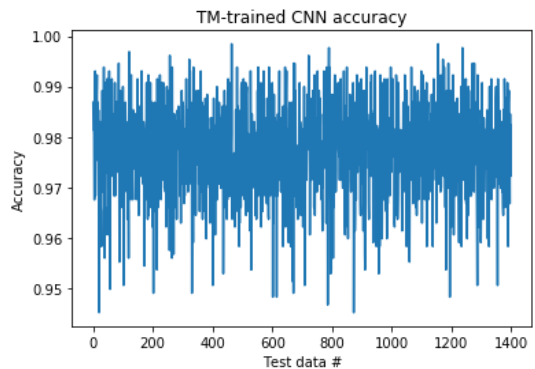

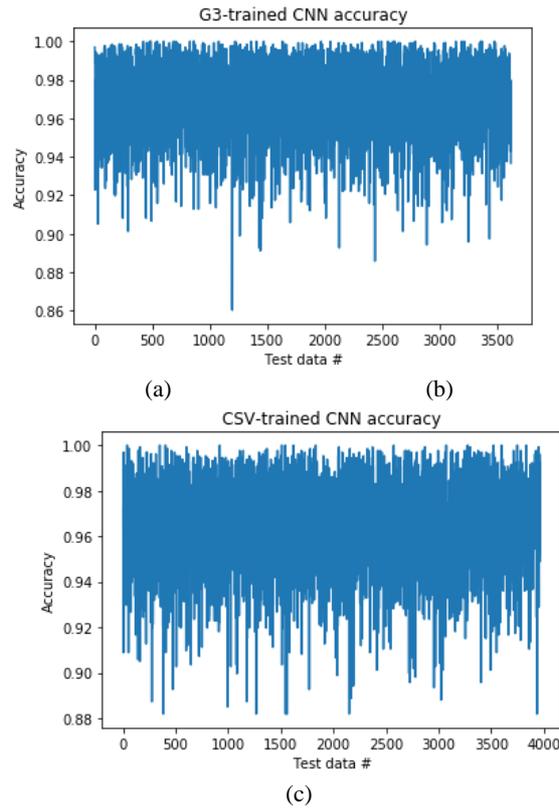

(a)                  (b)

(c)

Fig. S13 Prediction ccuracy distribution of CNNs over different datasets. (a) TM-trained CNN corresponding to Fig. 1(d) in the Main Text; (b) G3-trained CNN corresponding to Fig. 2(b) in the Main Text; (c) CSV-trained CNN corresponding to Fig. 3(b) in the Main Text.

Table S3. Average accuracy, f1_score, MSE, Corr, SSIM and output runtime using laptop CPU for different datasets.

| Models | Test set | Accuracy | F1_score | MSE | Corr | SSIM | Network output runtime (*sec*) |
|---|---|---|---|---|---|---|---|
| **TM-trained CNN** | 1,408 | 98.74% | 91.29% | 0.0167 | 0.9222 | 0.8478 | 0.0115 |
| **G3-trained CNN** | 3,622 | 97.06% | 86.95% | 0.0288 | 0.8621 | 0.7305 | 0.0117 |
| **G1&G2&G3-trained CNN** | 3,622 | 96.05% | 79.73% | 0.0348 | 0.7834 | 0.6644 | 0.0114 |
| **CSV-trained CNN** | 3,965 | 96.48% | 83.61% | 0.0313 | 0.8492 | 0.6968 | 0.0116 |

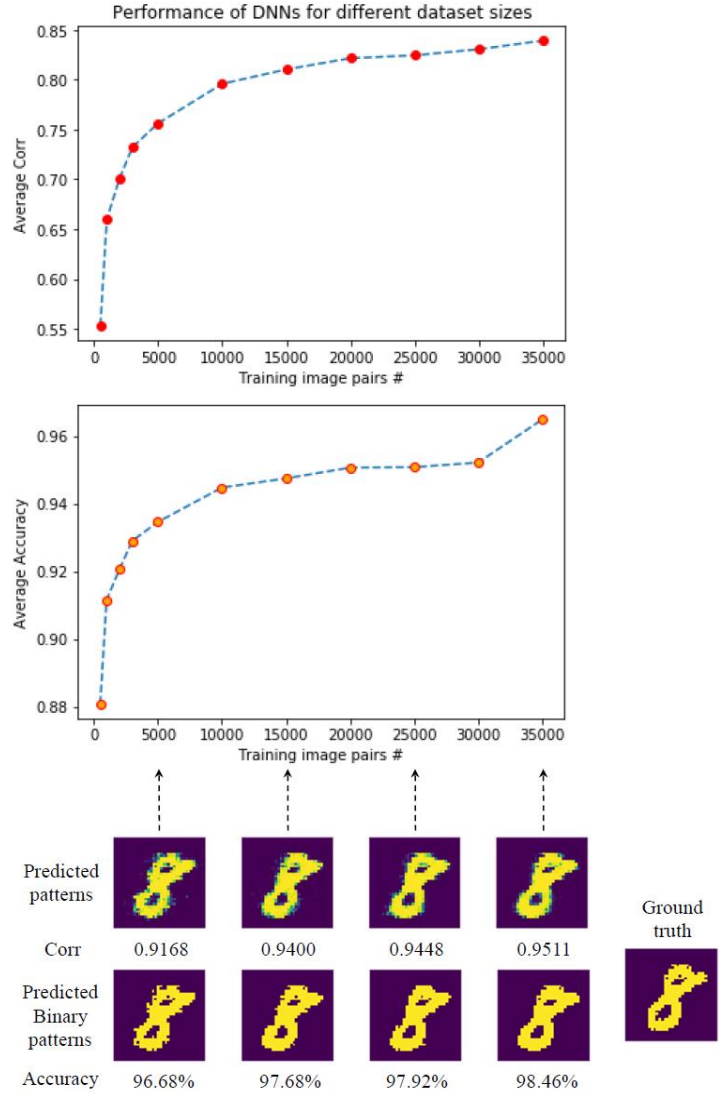

Fig. S14 The network performance (average prediction correlation coefficient and average prediction accuracy) at different sizes of randomly chosen training datasets (500, 1000, 2000, 3000, 5000, 10000, 15000, 20000, 25000, 30000 and 35000), tested using CSV-trained CNN and on associated testing dataset. The testing dataset size is 3,965. Insets below show the predicted patterns and their corresponding predicted patterns at different stages of the training.

Table S4. Training time of different dataset sizes

| Size of training dataset (including validation set) | Training time per epoch (*sec*) | Total training time (200 epoch) |
|---|---|---|
| 5000 patches | 14 | 47min |
| 10000 patches | 28 | 1hr,34min |
| 15000 patches | 46 | 2hr,33min |
| 20000 patches | 67 | 3hr,43min |
| 25000 patches | 80 | 4hr,27min |
| 30000 patches | 91 | 5hr,03min |

In order to understand the influence of the size of training dataset on the performance of deep neural networks, we randomly chose different sizes of datasets from the training dataset collected when the MMF is subject to CSV, as shown in Fig. S14. As expected, the increase of training data pairs results in stronger prediction ability.

**Correlation coefficient evaluations of different MMF transmission states**

Table S4. Correlation coefficient between the output speckle images with the same input DMD pattern at

steel bar locations: H1 to H6 when the MMF is subject to CSV

|    | H1   | H2   | H3   | H4   | H5   | H6   |
|----|------|------|------|------|------|------|
| H1 | 100% | 71%  | 60%  | 56%  | 51%  | 52%  |
| H2 | 71%  | 100% | 72%  | 56%  | 50%  | 52%  |
| H3 | 60%  | 72%  | 100% | 67%  | 50%  | 54%  |
| H4 | 56%  | 56%  | 67%  | 100% | 52%  | 53%  |
| H5 | 51%  | 50%  | 50%  | 52%  | 100% | 51%  |
| H6 | 52%  | 52%  | 54%  | 53%  | 51%  | 100% |

**Supplement Info References:**